\documentclass[twocolumn]{jpsj3}
\usepackage{txfonts,braket,color,ulem}

\title{Antiferromagnetic State in $\kappa$-type Molecular Conductors: \\
Spin Splitting and Mott Gap}

\author{Hitoshi Seo$^{1,2}$\thanks{seo@riken.jp} and Makoto Naka$^3$\thanks{naka@aoni.waseda.jp}}
\inst{$^1$Condensed Matter Theory Laboratory, RIKEN, Wako, Saitama 351-0198, Japan \\
$^2$RIKEN Center for Emergent Matter Science, Wako, Saitama 351-0198, Japan \\
$^3$Waseda Institute for Advanced Study, Waseda University, Shinjuku, Tokyo 169-8050, Japan} 

\abst{We numerically investigate the dynamical properties of $\kappa$-type molecular conductors in their antiferromagnetic Mott insulating state. 
By treating the extended Hubbard model on the two-dimensional $\kappa$-type lattice within the Lanzcos exact diagonalization method, 
we calculate the one-particle spectral function $A(\boldsymbol{k}, \omega)$ and the optical absorption spectra  
taking advantage of twisted boundary conditions. 
We find spin splitting in $A(\boldsymbol{k}, \omega)$ predicted by a previous Hartree--Fock study [M. Naka \textit{et al.}, Nat. Commun. {\bf 10}, 4305 (2019)];  
namely, their up-spin and down-spin components become different in the general $\boldsymbol{k}$-points 
of the Brillouin zone, even without the spin--orbit coupling.
Furthermore, we demonstrate the variation in the optical properties near the Mott gap caused by the presence or absence of the antiferromagnetic order,
tuned by a small staggered magnetic field. }

\begin{document}
\maketitle

\section{Introduction}

\begin{figure}
\vspace*{2em}
\begin{center}
\includegraphics[width=7cm,clip]{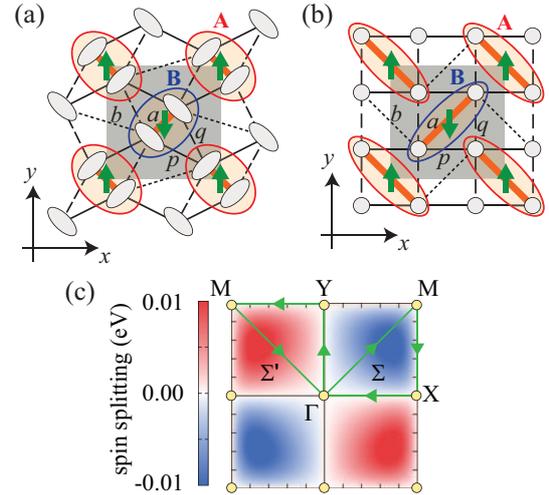}
\end{center}
\vspace*{-0.5em}
\caption{(Color online) (a) Schematic illustration of the $\kappa$-type lattice structure and (b) its deformed version on the square lattice. 
The ellipses in (a) and the circles in (b) represent the molecular sites, and the networks of the bonds, 
$a$~(bold line), $b$~(dotted line), $p$~(solid line), and $q$~(broken line), are shown. 
The grey squares show the unit cell common to the paramagnetic state and the AFM state,  
where two types of dimers $A$ and $B$ take opposite spin directions as shown in the figure by arrows. 
(c) Spin splitting in the first Brillouin zone within the Hartree--Fock approximation, for the Hubbard model with $U=1$ eV, predicted in Ref.~\citen{Naka_spincurrent}. 
The contour map shows the difference between the up-spin and down-spin energies of the top band (see Fig.~\ref{fig2}), 
and the trajectory shows the symmetric lines in Fig.~\ref{fig2}.}
\label{fig1}
\vspace*{-2.5em}
\end{figure}

$\kappa$-type molecular conductors, especially $\kappa$-(BEDT-TTF)$_2X$ where BEDT-TTF = bis(ethylenedithio)tetrathia-fulvalene and $X$ taking different anions, 
have been providing a fertile platform for strongly correlated electron physics\cite{Kanoda}. 
The intensively investigated phenomena include the Mott metal--insulator transition~\cite{Lefebre,Kagawa}, 
superconductivity~\cite{Urayama,Ardavan} and its high-field transition to inhomogeneous states~\cite{Ardavan,Brown}, antiferromagnetism~\cite{Miyagawa}, 
quantum spin-liquid behavior\cite{Shimizu}, and electronic dielectricity~\cite{Naka_electricferro,Ishihara,Hassan}. 
These rich and diverse properties, surprisingly, arise from a common framework: mutually interacting electrons on 
a two-dimensional arrangement of BEDT-TTF molecules, 
which we call here the $\kappa$-type lattice, as shown in Fig.~\ref{fig1}(a). 

Recently, Naka \textit{et al.} have proposed further possibilities of $\kappa$-type molecular conductors
within their antiferromagnetic (AFM) states: 
spin current transport~\cite{Naka_spincurrent} 
and anomalous Hall effect~\cite{Naka_AHE}. 
The former leads to a peculiar type of spin current generation when an electric field or a thermal gradient is applied to the system.
This mechanism does not rely on the atomic spin--orbit coupling as in the spin Hall effect, 
but is based on the interplay between the characteristic $\kappa$-type molecular arrangement and the collinear AFM order. 
Such an interplay leads to a spin-split band structure even without the spin--orbit coupling, which was overlooked in previous works 
on $\kappa$-(BEDT-TTF)$_2X$, 
and is now examined in a wide range of materials showing collinear-type AFM orderings
as well.~\cite{Noda,Okugawa,Hayami,Gonzalez,Yuan,Hayami2,Naka_perovskite}

For the realization of spin splitting, 
the importance of the $\kappa$-type molecular arrangement has been stressed~\cite{Naka_spincurrent}. 
Historically, the half-filled Hubbard model on the anisotropic triangular lattice has been intensively studied as an effective model of $\kappa$-(BEDT-TTF)$_2X$ 
and sucessfully applied to different aspects, especially the Mott transition~\cite{Kino,Furukawa}. 
This model is achieved by taking the molecular dimers [see Fig.~\ref{fig1}(a)] as lattice sites, which is called the strong dimerization limit.~\cite{Kino} 
However, the spin splitting does not occur within this model; therefore, maintaining the finite dimerization, 
namely, considering the frontier orbital for each molecule, is crucial for its realization and the resultant spin current transport. 
The effects of such an intradimer degree of freedom on the electronic properties in $\kappa$-type materials have been examined, 
mostly in direct relation to electric dielectricity and ferroelectricity~\cite{Naka_electricferro,Ishihara,Hassan}, 
and also to the superconducting state~\cite{Kuroki,Sekine,Guterding,Watanabe,Watanabe2}; 
here, we focus on its effect on the AFM state. 

In this paper, 
we investigate the spin splitting in $\kappa$-type molecular conductors, 
including the strong correlation effect by treating the system fully considering the quantum effects. 
The treatment in Ref.~\citen{Naka_spincurrent} was within the Hartree--Fock (HF) approximation to the Hubbard model on the $\kappa$-type lattice. 
However, such a treatment ignores fluctuations beyond mean field and therefore 
cannot correctly describe the Mott physics observed experimentally. 
Here, our aim is to investigate the realization of spin splitting quantitatively 
in the Mott insulating regime beyond the mean field. 
By treating the extended Hubbard model (EHM) on the $\kappa$-type lattice within the exact diagonalization method, 
we indeed find the spin-split structure in the one-particle spectral function. 
We further study the optical absorption spectra, which can be used to directly probe the Mott gap,  
and the effect of the AFM order on them. 

\section{Model and Method}

First, let us introduce the model and method. 
Notably, we use twisted boundary conditions for the treatment of a finite-size cluster, 
to make a continuous variation of $\boldsymbol{k}$-points in the Brillouin zone for the one-particle spectral function 
and to reliably evaluate the optical spectra by taking the average for many boundary conditions.
Our model is the EHM based on the frontier orbital for each molecule~\cite{Kino,Seo_ET,Seo_ChemRev}, 
whose Hamiltonian is described as 
\begin{align}
{\cal H}_0 =\sum_{\langle i,j \rangle, s} t_{ij} \left( c^\dagger_{is} c_{js}^{} + \mathrm{h.c.} \right)
 + U \sum_{i} n_{i\uparrow} n_{i\downarrow} 
 + \sum_{\langle i,j \rangle} V_{ij} n_{i} n_{j},\label{eq1}
\end{align}
where $c_{is}$ ($c^\dagger_{is}$) and $n_{is}$ ($= c^\dagger_{is}c_{is}$) are the annihilation (creation)
and number operators of an electron at the $i$th molecular site 
with spin $s$, respectively, and $n_i =  n_{i\uparrow}+ n_{i\downarrow}$. 
$t_{ij}$ and $V_{ij}$ represent the intersite
transfer integrals and Coulomb interactions on the site pairs $\langle i,j \rangle$, 
and $U$ is the intrasite Coulomb interaction. 
We consider the pairs along the bonds shown in Fig.~\ref{fig1}(a): the intradimer $a$ and the interdimer $p, q$, and $b$ bonds. 
A deformed network on the square lattice is drawn in Fig.~\ref{fig1}(b), 
which can be viewed as a modified Shastry--Sutherland lattice~\cite{Shastry}. 
The lattice structures in Figs.~\ref{fig1}(a) and~\ref{fig1}(b) are geometrically equivalent, 
but for the optical responses written below, 
the lattice vectors, which are material-dependent, appear in the expression; 
here, we use those on the square lattice shown in Fig.~\ref{fig1}(b), aiming for a general investigation.  

We set the transfer integrals to the parameters obtained by a first-principles band calculation for 
$\kappa$-(BEDT-TTF)$_2$Cu[N(CN)$_2$]Cl as $(t_a, t_p, t_q, t_b) = (-0.207, -0.102, 0.043, -0.067)$~eV.~\cite{Koretsune}
In this parameter range, at three-quarter filling, 
 an AFM order with the pattern shown in Figs.~\ref{fig1}(a) and \ref{fig1}(b) 
 is stabilized under strong correlation~\cite{Kino,Watanabe}. 
However, in the finite size cluster calculation, which we describe below, 
 the ground state becomes a linear combination between the doubly degenerate AFM states, resulting in a nonmagnetic singlet state. 
To extract one of the AFM patterns, we add a dimer-dependent magnetic field coupled to the collinear AFM state 
as an external field term described by 
\begin{align}
{\cal H}_\mathrm{ext} = - \ h_\mathrm{AFM} \left\{ \sum_{i \in A} \left(n_{i\uparrow} - n_{i\downarrow}\right) - 
\sum_{i \in B} \left(n_{i\uparrow} - n_{i\downarrow}\right) \right\}, 
\end{align}
where $h_\mathrm{AFM}$ is the AFM field, 
and $A$ and $B$ represent the different dimers [see Figs.~\ref{fig1}(a) and \ref{fig1}(b)]. 

To investigate the dynamical properties %
taking into account the strong correlation effect properly, 
we use the Lanczos exact diagonalization technique. 
By taking advantage of the standard methods using the continued fraction expansion~\cite{Dagotto}, 
one can compute the spin-dependent one-particle spectral function $A_s(\boldsymbol{k}, \omega)$ as 
\begin{align}
A_s(\boldsymbol{k}, \omega) = A^{\mathrm{el}}_{s}(\boldsymbol{k}, \omega) + A^{\mathrm{h}}_{s}(\boldsymbol{k}, \omega), 
\end{align}
where $A^{\mathrm{el}}_{s}(\boldsymbol{k}, \omega)$ and $A^{\mathrm{h}}_{s}(\boldsymbol{k}, \omega)$ are 
the electron-creation and electron-annihilation (i.e, hole-creation) spectral functions with spin $s$, respectively. 
The former is obtained as 
\begin{align}
A^{\mathrm{el}}_{s}(\boldsymbol{k}, \omega) =  \frac{1}{\pi} \sum_{l} \mathrm{Im} 
\Braket{0  |  
c_{l\boldsymbol{k}s} \frac{1}{\omega-{\cal H} + E_{0} -\mu -i\epsilon} c_{l\boldsymbol{k}s}^\dagger
| 0 }, 
\end{align}
where, 
by noting that there are four sublattices in the unit cell, 
$c_{l\boldsymbol{k}s}$ ($c_{l\boldsymbol{k}s}^\dagger$) is the Fourier transform of  $c_{is}$ ($c_{is}^\dagger$) 
with $i \in l$, with $l$ ($= 1$ -- $4$) being the sublattice number in the unit cell. 
Namely, 
\begin{align}
c_{l\boldsymbol{k}s} = \frac{4}{N} \sum_{i\ \in \ l} e^{i \boldsymbol{r}_i \cdot \boldsymbol{k}} \ c_{is} \ ; 
\ c_{l\boldsymbol{k}s}^\dagger = \frac{4}{N} \sum_{i\ \in \ l} e^{- i \boldsymbol{r}_i \cdot \boldsymbol{k} } \ c_{is}^\dagger, 
\end{align}
where $N$ and $\boldsymbol{r}_i$ are the system size (total number of unit cells is therefore $N/4$) and the position of the $i$th site, respectively.  
$\ket{0}$ and $E_{0}$ are the ground-state eigenvector and its energy, respectively; 
$\mu$ is the chemical potential defined as half of the difference between the ground-state energies of the one-electron-added and one-electron-subtracted systems. 
The hole-creation spectral function $A^{\mathrm{h}}_{s}(\boldsymbol{k}, \omega)$ is given, on the rhs of Eq.~(4) by  
exchanging $c_{l\boldsymbol{k}s}$ and $c_{l\boldsymbol{k}s}^\dagger$, 
and replacing $-{\cal H} + E_{0}$ with $+{\cal H} - E_{0}$ in the denominator. 

\begin{figure*}[t]
\vspace*{1em}
\begin{center}
\includegraphics[width=17cm,clip]{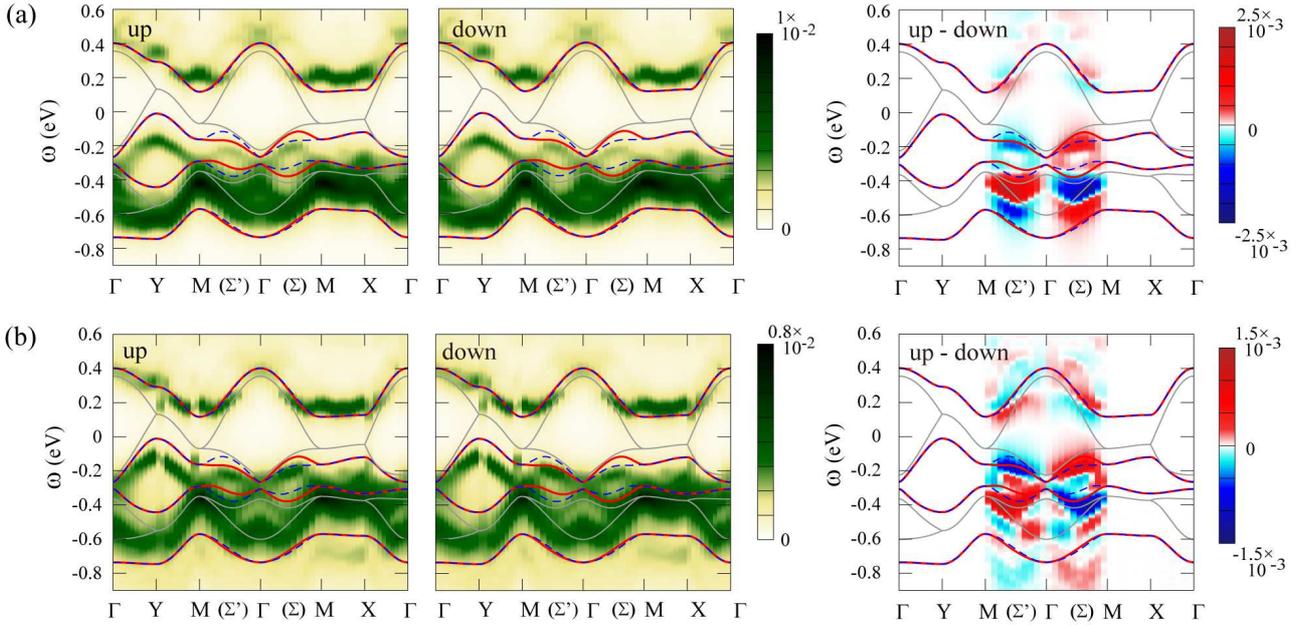}
\end{center}
\vspace*{-0.5em}
\caption{(Color) Spin-dependent one-particle spectral functions $A_{s}(\boldsymbol{k}, \omega)$ ($s = \uparrow$, $\downarrow$) for the two parameter sets
 (a) $t_a \rightarrow 2t_a$, $U=5$~eV, $V_{ij}=0$ and (b) $U=2$~eV, $(V_a, V_p, V_q, V_b) = (0.5, 0.3, 0.3, 0.3)$~eV, 
under the AFM field $h_\mathrm{AFM} = 0.01$~eV.
Left, middle, and right panels show the up- and down-spin components, and their differences 
$A_{\uparrow}(\boldsymbol{k}, \omega) - A_{\downarrow}(\boldsymbol{k}, \omega)$, respectively; 
we set the broadening as $\epsilon = 0.05$ eV, and the Fermi level is $\omega=0$. In each cut of the Brillouin zone, we take 10 meshes. 
The grey solid lines are the tight-binding band structure ($U=0$), 
and the red solid and blue broken 
lines are the up- and down-spin bands, respectively, in the AFM state
 within the Hartree--Fock approximation for $U=1$~eV, $V_{ij}=0$~\cite{Naka_spincurrent}.}
\label{fig2}
\vspace*{-2em}
\end{figure*}

The drawback of this framework is that one cannot make the system size $N$ large owing to computational limitations. 
In the results that follow, we use a $N=4 \times 4 = 16$-site lattice: $2 \times 2$ of the unit cell; 
the number of electrons is fixed at three-quarter filling, i.e., $3N/4 \times 2 = 24$.   
In such a cluster, within the periodic boundary condition, the finite $\boldsymbol{k}$-points are only allowed at the high-symmetry points, 
but the spin splitting is predicted to appear only in general $\boldsymbol{k}$-points in the Brillouin zone, 
as shown in Fig.~\ref{fig1}(c). 
Therefore, we use the twisted boundary condition, which makes the possible $\boldsymbol{k}$-points continuous, 
as developed in Refs.~\citen{Poilblanc,Tsutsui,Tohyama}. 
In this case, $\ket{0}$, $E_{0}$, and $\mu$ are calculated within each choice of the twisted boundary condition so as to fit the target $\boldsymbol{k}$-point. 

We also consider the twisted boundary condition in calculating the optical absorption spectra 
defined as 
\begin{align}
\alpha_\nu(\omega) = - \frac{e^2}{N} \mathrm{Im}  \Braket{0  |  
j_\nu \frac{1}{\omega-{\cal H} + E_{0} -i\epsilon} j_\nu 
| 0 }, 
\end{align}
where $\nu$ is the direction of light polarization in the $xy$ plane and $j_\nu$ is the $\nu$ component of the electric current operator defined as 
\begin{align}
\boldsymbol{j} = i \sum_{\langle i,j \rangle, s} t_{ij} \left(\boldsymbol{r}_i-\boldsymbol{r}_j \right) 
\left( c^\dagger_{is} c_{js}^{} - \mathrm{h.c.} \right). 
\end{align}
Here, for the position of the lattice sites $\boldsymbol{r}_i$, 
we take the square lattice geometry in Fig.~\ref{fig1}(b) and set half of the lattice constants to 1 for both $x$ and $y$ directions. 
The relation $\alpha_\nu(\omega) = \omega \ \sigma_\nu(\omega)$ holds, 
where $\sigma_\nu(\omega)$ is the regular part of the optical conductivity spectra, without the Drude term.  
By taking an average on the twisted boundary conditions, we can provide results with a lessened dependence on the boundary condition.~\cite{Tohyama} 

\section{Results}

Before presenting the results of numerical calculations, 
let us briefly review the electronic structure within the HF approximation~\cite{Naka_spincurrent}.
The HF band structure of the AFM insulating state is given in Fig.~\ref{fig2}, 
 with up spin (red solid lines) and down spin (blue broken lines), 
for the Hubbard model with $U=1$~eV. 
For this choice of path in the Brillouin zone, 
 the two dispersions take different values along the M-($\Sigma$')-$\Gamma$-($\Sigma$)-M line, 
 while along the other lines, they are degenerate [see Fig.~\ref{fig1}(c)].

This spin splitting, as mentioned above, is not due to the spin--orbit coupling, as it is not included in the model, 
 but originates from the collinear AFM pattern within the $\kappa$-type lattice structure. 
From the symmetry point of view, 
 the AFM pattern globally breaks the time-inversion symmetry; 
 namely, starting from one of the doubly degenerate AFM states, 
 the other state is not connected to the original one by flipping the spin directions and a further translation or inversion operation, 
 but is connected by a glide operation. 
This makes the lifting of the Kramers degeneracy possible, which is in clear contrast to simpler AFM states (the N\'{e}el state), 
 for example, on the square lattice.

An intuitive picture is given as follows~\cite{Naka_spincurrent}, 
 by considering the effective hopping integrals for the $A$ and $B$ dimers hosting opposite spin directions. 
Since the two dimers have different orientations, when the AFM spin pattern shown in Fig.~\ref{fig1}(a) is stabilized, 
 the up (down) spins on $A$ ($B$) dimers have the tendency to hop more along the [11] ([$\overline{1}$1]) direction. 
This results in the spin-dependent energy dispersion along such directions, as demonstrated in Fig.~\ref{fig1}(c).

In Ref.~\citen{Naka_spincurrent}, the Hubbard model on the $\kappa$-type lattice was considered 
using the same values for $t_{ij}$ as in the present study, but the $V_{ij}$ terms were not included. 
We consider the EHM for a more qualitative description; 
in fact, when using numerical methods considering the quantum fluctuations, 
the inclusion of the $V_{ij}$ terms is crucial for the stabilization of the Mott insulating state~\cite{Watanabe,Gomi,Sato}, 
called the dimer Mott insulator. 
On the other hand, it can be stabilized by artificially making $t_a$ large; 
in this case, the Hubbard $U$ term suffices for its stabilization, without the $V_{ij}$ terms. 
Therefore, in the following, we use two parameter sets for investigating the typical behavior of the dimer Mott insulator in the $\kappa$-type lattice structure: 
(a) modifying the intradimer transfer integral from the value above as $ t_a\rightarrow 2t_a = -0.414$~eV, and setting $U=5$~eV, $V_{ij}=0$, 
(b) using the first-principles values above for $t_{ij}$, and $U=2$, $(V_a, V_p, V_q, V_b) = (0.5, 0.3, 0.3, 0.3)$~eV, referring to previous works,~\cite{Watanabe,Mori,Naka_excitation,Naka_cat} 
which is considered to be closer to the experimental situation.

{We use the AFM field of $h_\mathrm{AFM} =$ 0.01~eV which is larger than the spin gap in our finite size cluster, 
 needed for the stabilization of the AFM state. 
In this case, the magnetic moment on each site becomes about 0.4~$\mu_{\rm{B}}$ (0.8~$\mu_{\rm{B}}$ per dimer), 
close to the experimental value~\cite{Miyagawa} 
as well as to the HF results for the Hubbard model at $U=$ 1~eV~\cite{Kino,Naka_spincurrent}. 
The results below do not change qualitatively in the parameter range (typically,  $h_\mathrm{AFM} > 0.005$~eV), where such an AFM order is stabilized.

\subsection{One-particle spectral function}

In Fig.~\ref{fig2}, we show the spin-dependent one-particle spectral functions $A_{s}(\boldsymbol{k}, \omega)$ for the two parameter sets (a) and (b), 
for the up-spin (left panels) and down-spin (middle panels) components and their differences 
$A_{\uparrow}(\boldsymbol{k}, \omega) - A_{\downarrow}(\boldsymbol{k}, \omega)$ (right panels); 
a broadening of $\epsilon = 0.05$ eV is used. 
The spectral functions show that the system is in the Mott insulating regime; a gap opening is seen at the Fermi level for both parameter sets. 
Their basic structures are common to the nonmagnetic case with $h_\mathrm{AFM}=0$~\cite{supplement}, 
whose upper two bands show similarities to the half-filled Hubbard model,~\cite{Kohno,Kawasugi} which is valid in the strong dimerization limit. 
Namely, the doubly degenerate bands in the noninteracting case, which cross the Fermi energy, along the zone boundary split into two, and a gap is opened. 
This is similar to the HF results, but one different aspect is that, 
in the HF approximation, the whole band structure is affected by the AFM order since it is included as a mean field, 
while, in the results here, the lower two bands far from the Fermi level are insensitive. 
This is a result characteristic of strong correlation; the structure near the Fermi level opens the gap resulting in the Mott insulating state~\cite{Georges}. 

The effect of the external field $h_\mathrm{AFM}$ can be observed in the asymmetric structures 
with respect to the $\Gamma$-point 
along the M-($\Sigma$')-$\Gamma$-($\Sigma$)-M line. 
The spin splitting 
is clearly demonstrated in the spin differences, 
while the overall structure is not considerably affected. 
Similarly to the HF prediction, the spin splitting appears only in the general $\boldsymbol{k}$-points and is largely seen in the area below the Fermi level;  
in fact, the contrast exists in a broad range of energy well below the Fermi level.
The sign of spin splitting just below the Fermi level is consistent with the HF results, 
especially for the parameter set (a); for the parameter set (b), a more complex behavior is seen.  
The sign is, however, opposite to that in the HF results [Fig.~\ref{fig1}(c)] in the area above the Fermi level; 
along the $\Gamma$-($\Sigma$)-M [M-($\Sigma'$)-$\Gamma$] line, a spin-down (spin-up) band is predicted to be higher in energy in HF,  
whereas the spin-up (spin-down) component is larger in the higher energy region of the top band in our results here.

\subsection{Optical spectra}

Next, we show the optical absorption spectra $\alpha_\nu(\omega)$ for the two parameter sets in Figs.~\ref{fig3}(a) and \ref{fig3}(b). 
The twisted boundary condition average is taken for 20 $\times$ 20 meshes in the Brillouin zone. 
The broadening is set to $\epsilon=0.005$~eV, narrower than the value used above, to see the fine structures. 
We also plot the density of states for comparison, defined as the $\boldsymbol{k}$-sum of $A_s(\boldsymbol{k}, \omega)$, with the same $\epsilon$ and meshes. 
Note that the density of states is spin-independent even for the spin-split case through the $\boldsymbol{k}$-sum; 
the spin splitting has the inversion symmetry in two-dimensional $\boldsymbol{k}$-space as in the HF results. 
In Fig.~\ref{fig3}, the results for both polarizations $\nu=x,y$ are shown together with the density of states, with and without the AFM order 
tuned by switching on and off the external AFM field $h_\mathrm{AFM}$. 

\begin{figure}
\vspace*{1em}
\begin{center}
\includegraphics[width=7cm,clip]{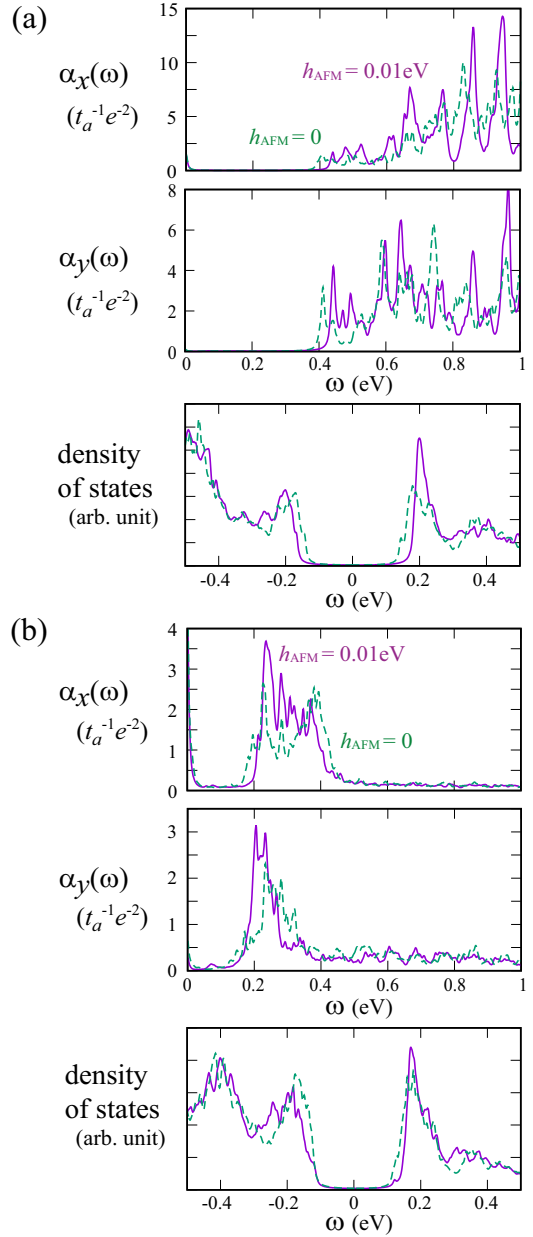}
\end{center}
\vspace*{-0.5em}
\caption{(Color online) Optical absorption spectra $\alpha_\nu(\omega)$ for the two polarization directions $\nu=x$ and $y$ and the density of states 
for the two parameter sets (a) $t_a \rightarrow 2t_a$, $U=5$~eV, $V_{ij}=0$ and (b) $U=2$~eV, $(V_a, V_p, V_q, V_b) = (0.5, 0.3, 0.3, 0.3)$~eV, 
with and without the AFM field $h_\mathrm{AFM} = 0.01$~eV.
A twisted boundary condition average is taken for 20 $\times$ 20 meshes with a broadening factor $\epsilon=0.005$~eV.}
\vspace*{-2em}
\label{fig3}
\end{figure}

One can see the appearance of the gap in the optical responses in both parameter sets. 
Note that the features near $\omega=$~0 contain artifacts due to the twisted boundary conditions; 
 the enhancements toward $\omega=$~0 seen in $\alpha_x(\omega)$ are absent in the periodic/antiperiodic boundary conditions and there is a clear gap, 
 but they appear in some choices of twisted boundaries. 
This is presumably attributable to the discreteness of the $\boldsymbol{k}$-points (the shell problem)
 resulting in the chemical potential to be off the gap structure, 
 which is also reflected in Fig.~\ref{fig2} as discrete jumps in the dispersion relation $A(\boldsymbol{k}, \omega)$.
The gap evaluated from the density of states is approximately 0.35 and 0.25~eV for (a) and (b), respectively;  
 it is an indirect gap as seen in Fig.~\ref{fig2}, 
 where the band bottom above the Fermi energy is near the M-X line, whereas the band top below the Fermi energy is around the Y point in both cases

In the dimer Mott insulating states, 
 it has been discussed that the low-energy optical transitions across the gap 
 can be classified into two contributions~\cite{Faltermeier,Hashimoto,Ferber}. 
One is the intradimer excitation, which is a transition from the bonding to the antibonding bands, 
 and the other is the so-called Hubbard excitation, which corresponds to the transition from the lower to the upper Hubbard bands.
Their excitation energies can be roughly estimated from the energy levels of the isolated dimers~\cite{Naka_excitation,Naka_cat}. 
The former is the bonding--antibonding splitting $2|t_a|$ 
  and  the latter corresponds to the effective intradimer Coulomb energy 
  $U_\textrm{eff} = 2|t_a| + (U + V_a)/2 - \sqrt{4t_a^2 + (U - V_a)^2/4}$~\cite{Tamura}.

For the parameter set (a), these estimates give $2|t_a| = 0.83$ eV and $U_\textrm{eff} = 0.69$ eV. 
Therefore, the lowest optical transition across the gap is presumably the Hubbard excitation. 
The lowest excitation energy seen in Fig.~\ref{fig3}(a) is about 0.4~eV; 
 the difference from $U_\textrm{eff} = 0.69$ eV is attributable to the band effect, 
 namely, the finite dispersion from the interdimer hoppings of the order of 0.1~eV. 

On the other hand, in the parameter set (b), one obtains $2|t_a|=$ 0.41~eV and $U_\textrm{eff} =$ 0.81~eV; 
these values indicate that the low-energy region originates from the intradimer excitation,  
while the broad band at higher energy $\omega > 0.4$ eV is from the Hubbard excitation. 
In fact, such an assignment is consistent with a study based on first principles~\cite{Ferber}, 
that is, by a combination of density functional theory calculations and the dynamical mean field theory approach to $\kappa$-(BEDT-TTF)$_2$Cu[N(CN)$_2$]Br$_x$Cl$_{1-x}$. 
The low-energy peak like structure with large spectral weight in the intradimer part and its notable polarization direction dependence
have been revealed as the tendency toward the polar charge ordered phase~\cite{Naka_excitation,Gomi}: 
a collective mode of intradimer charge order producing electric polarization.
This is why the excitations are seen well below $2|t_a| = 0.41$ eV.  
The lower energy in the $y$ direction in the lowest excitations at around 0.15~eV than in the $x$ direction around 0.2~eV is 
 owing to the softening toward 
the so-called ferroelectric charge order phase with polarization along the $y$ axis.

The effect of $h_\mathrm{AFM}$, first, in all the results here, is to increase the gap, clearly seen in the density of states. 
In the absorption spectra, the small AFM field makes the edge of the gap structure 
slightly shifted to larger $\omega$ in both (a) and (b) compared with the $h_\mathrm{AFM}=0$ case. 
A large transfer of spectral weight is prominently seen in (b), 
where a large increase in absorption by the inclusion of $h_\mathrm{AFM}$ 
is seen in the region $\omega \sim 0.25$~eV in $\alpha_x(\omega)$ and $\omega \sim 0.2$~eV in  $\alpha_y(\omega)$. 
These are most probably related to the tendency toward charge order, although we leave the detailed investigation to future works. 
Indeed, the coupling between the magnetic and charge sectors in the $\kappa$-type materials 
has been discussed~\cite{Naka_electricferro,Gomi,Naka_excitation,Naka_multiferro,Li,Hotta,Lunkenheimer,Tomic}.   
Considering that the choice of parameter set (b) is closer to the experimental situation, 
it will be interesting to carefully analyze the temperature dependence of optical responses near the AFM ordering temperature.

\section{Discussion and Summary}

Let us briefly discuss the role of three dimensionality, 
namely, the interlayer coupling, which is weak and therefore neglected in this study. 
Nevertheless, it is discussed that the coupling is important in a sense that it determines the stacking patterns of the AFM order, 
which is directly related to the observation of anomalous transport properties. 
In $\kappa$-(BEDT-TTF)$_2$Cu[N(CN)$_2$]Cl and deuterated $\kappa$-(BEDT-TTF)$_2$Cu[N(CN)$_2$]Br, 
it has recently been revealed that their intralayer AFM patterns are common, as discussed in this work [see Fig.~\ref{fig1}(a)],
but the stacking patterns are different~\cite{Ishikawa,Oinuma}. 
There, the spin current conductivity is predicted to appear in the former but to become zero in the latter by cancellation, 
whereas the opposite behavior is predicted for the anomalous Hall effect~\cite{Naka_spincurrent,Naka_AHE}. 
The effect of the interlayer coupling on the spin splitting and its relationship with the anomalous transport are still to be fully understood 
and left as a future work. 

A related problem is the role of spin--orbit coupling in this system. 
Recently, there have been several theoretical works investigating how to incorporate the spin--orbit coupling into 
the (extended) Hubbard models for $\kappa$-(BEDT-TTF)$_2X$ 
and their effects on the electronic properties.~\cite{Winter,Jacko,Naka_AHE}
The interplay between the strong correlation effect studied in this work 
 and the spin--orbit coupling is another topic yet to be explored. 

In summary, 
we have investigated the dynamical properties of the dimer Mott insulating state showing antiferromagnetic ordering, 
within the extended Hubbard model on the $\kappa$-type lattice structure 
using the numerical exact diagonalization, fully taking into account the strong correlation effect.  
We find a spin-split structure in the one-particle spectral function in the collinear antiferromagnetic ordered state, 
consistent with the Hartree--Fock prediction in Ref.~\citen{Naka_spincurrent}. 
In the optical spectra, 
the antiferromagnetic ordering induces a shift in the gap toward higher energies than 
the nonmagnetic case with a transfer of spectral weight. 

\
 
\acknowledgment

We thank K. Tsutsui for fruitful discussions. 
This work was supported by JSPS KAKENHI Grant Numbers 
19K03723, 
19H01833, 
19K21860, 
20H04463, 
and 20H00121, 
and the GIMRT Program of the Institute for Materials Research, Tohoku University, No. 19K0019.

\end{document}